\documentclass[aps,prb,twocolumn,showpacs,groupedaddress]{revtex4}
\usepackage{here}
\usepackage{graphicx}
\begin{document}


\title{Determining ethylene group disorder levels in $\kappa$-(BEDT-TTF)$_2$Cu[N(CN)$_2$]Br}
\author{A.~U.~B. Wolter$^{1,2}$, R. Feyerherm$^2$, E. Dudzik$^2$, S. S\"{u}llow$^1$, Ch. Strack$^3$, M. Lang$^3$, D. Schweitzer$^4$}
\address{$^1$Institut f\"{u}r Physik der Kondensierten Materie, TU Braunschweig, 38106 Braunschweig, Germany\\
$^2$Hahn--Meitner--Institut GmbH, c/o BESSY, 12489 Berlin, Germany\\
$^3$Physikalisches Institut, J.W. Goethe Universit\"{a}t, FOR412, 60438 Frankfurt am Main, Germany\\
$^4$Physikalisches Institut, Universit\"{a}t Stuttgart, 70550 Stuttgart, Germany}

\date{\today}

\begin{abstract}
We present a detailed structural investigation of the organic superconductor $\kappa$-(BEDT-TTF)$_2$Cu[N(CN)$_2$]Br at temperatures $T$ from 9
to 300 K. Anomalies in the $T$ dependence of the lattice parameters are associated with a glass-like transition previously reported at $T_g$ =
77 K. From structure refinements at 9, 100 and 300 K, the orthorhombic crystalline symmetry, space group {\it Pnma}, is established at all
temperatures. Further, we extract the $T$ dependence of the occupation factor of the eclipsed conformation of the terminal ethylene groups of
the BEDT-TTF molecule. At 300 K, we find 67(2) \%, with an increase to 97(3) \% at 9 K. We conclude that the glass-like transition is not
primarily caused by configurational freezing-out of the ethylene groups.
\end{abstract}


\maketitle

Due to their exotic superconducting- and normal-state properties, which resemble those of the high-$T_c$ cuprates, the organic charge-transfer
salts $\kappa$-(BEDT-TTF)$_2$X, X = Cu(NCS)$_2$, Cu[N(CN)$_2$]Br and Cu[N(CN)$_2$]Cl have been intensively studied in recent
years~\cite{lang96,mckenzie97,lefebvre00,singleton02,muller02,strack05}. Resulting from their layered crystal structure, the electronic
properties of these materials are quasi-twodimensional. This leads to electronic ground state properties for the series of materials which are
commonly summarized within a conceptual phase diagram, with the antiferromagnetic insulator X = Cu[N(CN)$_2$]Cl and the correlated metals X =
Cu[N(CN)$_2$]Br, Cu(NCS)$_2$ on opposite sites of a bandwidth-controlled Mott transition~\cite{kanoda97}.

\begin{figure}
\begin{center}
\includegraphics[width=1\columnwidth]{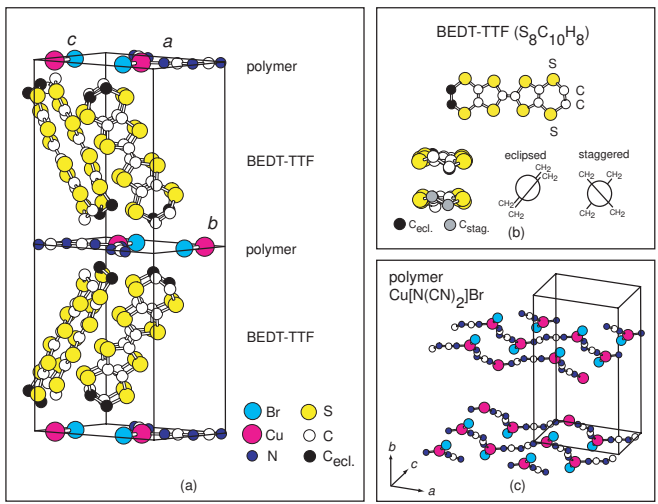}
\end{center}
\caption[1]{(Color online) (a) The crystallographic structure of $\kappa$-Br. (b) The {\it eclipsed} and {\it staggered} configurations of the
ethylene groups (CH$_2$)$_2$ of the BEDT-TTF molecule. Lower part shows the view along the long axis of the molecule. (c) View of the packing of
the polymeric anion chains in $\kappa$-Br. Protons on the BEDT-TTF molecule are omitted for clarity.} \label{fig:struc}
\end{figure}

In recent thermal expansion measurements on {\mbox{$\kappa$-(BEDT-TTF)$_2$Cu[N(CN)$_2$]Br}} (abbreviated $\kappa$-Br hereafter) three kinds of
anomalies have been identified~\cite{muller02}: one associated to the superconducting transition at $T_c = 11.6$~K, a phase-transition-like
anomaly at $T^\star \approx$~45~K, and a kinetic glass-like transition at $T_g = 77$~K. The origin of the $T^\star$ anomaly, coinciding with
distinct features in bulk properties, is still unclear~\cite{lang96}. Likewise, the nature of the transition at $T_g$ remains enigmatic. The
lattice response reveals the characteristics of a glass-like transition~\cite{muller02}, where below $T_g$ some short-range order is frozen in.
From the various hypotheses~\cite{muller02,su98,tanatar99,sasaki05}, the idea of a configurational freezing-out of the terminal ethylene groups
of the BEDT-TTF molecule at $T_g$ received most support.

Fig.~\ref{fig:struc} displays the orthorhombic crystal structure of $\kappa$-Br, space group {\it Pnma} with $Z$ = 4. It contains alternating
layers of a conducting BEDT-TTF cation network (S$_8$C$_{10}$H$_8$)$_2$$^+$ (Fig.~\ref{fig:struc}(b)) and insulating polymeric anion chains
Cu[N(CN)$_2$]Br$^-$ (Fig.~\ref{fig:struc}(c)) along the $a$ direction and shifted by one quarter of a unit cell along $b$. The BEDT-TTF molecule
network contains centrosymmetric face-to-face dimers in a typical $\kappa$-type packing motif. Here, the $\pi$-framework of each BEDT-TTF donor
molecule is inclined to the anion layer, so the ethylene groups (CH$_2$)$_2$ of the donor molecules realize short
{\mbox{C--H$\cdot\cdot\cdot$anion}} contacts. The polymeric anion, which resides on a mirror plane, contains infinite zig-zag chains of
[Cu--dicyanamide--Cu] units, and the copper atoms complete their coordination spheres with the Br atom.

In addition, Fig.~\ref{fig:struc}(b) schematically depicts the two possible relative orientations of the terminal ethylene groups (CH$_2$)$_2$
of the BEDT-TTF molecule. While the outer C--C bonds are parallel in the {\it eclipsed} configuration, they are canted against each other in the
{\it staggered} one. At high temperatures the terminal groups are thermally excited and fluctuate between the two states, but upon lowering the
temperature a preferential orientation in the eclipsed configuration occurs. It has been suggested that at $T_g$ these excitations freeze out,
giving rise to the glass-like phenomena~\cite{muller02}. Then, the ''glassy'' state below $T_g$ would be characterized by the level of frozen-in
disorder, which will strongly depend on the cooling rate employed experimentally. Such residual disorder might also be the cause for
controversial experimental findings concerning the symmetry of the superconducting order parameter of these materials. It might also account for
the pronounced cooling-rate dependence of various normal state properties such as the resistivity~\cite{tanatar99,stalcup99}. Hence, for a full
understanding of these materials a detailed knowledge of the structural properties is essential.

Despite extensive experimental efforts to unravel the nature of the anomalous glass-like transition at $T_g$, a direct proof for a relation
between a configurational freezing-out of the ethylene groups, with associated residual ethylene group disorder, and the glass-like transition
is lacking, although the accompanied activation energy~\cite{muller02,wzietek96} and the observed isotope effect~\cite{muller02} indicate that
the terminal ethylenes are involved. While Refs.~\cite{muller02,akutsu00} present ample evidence for the existence of a glass-like structural
anomaly, the crystallographic structure at low temperatures $T$ has never been determined with very high resolution. In fact, x-ray diffraction
studies~\cite{geiser91,geiser91a} carried out at 127 and 20 K claimed the observation of a fully ordered structure at these temperatures, with
the ethylene groups exclusively adopting the eclipsed configuration (within an experimental uncertainty~\cite{muller02} of $\sim$ 10 \%). These
studies imply that a disorder-order transition of the terminal (CH$_2$)$_2$ groups should occur above 127 K. Structural investigations by
Watanabe et al.~\cite{watanabe91} identified an irregular $T$ dependence of the structural parameters near $T_g$, but were unable to associate
this to a disorder-order transition of the terminal ethylene groups.

In this Communication we present a single crystal structural investigation of the organic superconductor $\kappa$-Br at temperatures 9 K $\leq T
\leq$ 300 K combining laboratory structure refinements and high-resolution synchrotron x-ray diffraction experiments. The single crystal studied
here was grown using a non-standard preparation route. It has been characterized in full detail including resistivity
experiments~\cite{strack05} (sample $\sharp$3 with a residual resistivity ratio of 84 and a low resistive behavior around 90~K, which contrasts
with ''high resistive'' behavior observed for materials prepared along the standard route; see Ref.~\cite{strack05} for a detailed discussion).
Distinct anomalies are identified in the $T$ dependence of all three lattice parameters. Structure refinements between 9 and 300 K reveal that
the crystalline symmetry remains orthorhombic, space group {\it{Pnma}} ($Z$ = 4). We determine the residual level of orientational disorder of
ethylene configurations to 3$\pm$3 \% at lowest temperatures.

Synchrotron x-ray diffraction experiments were performed on the 7~T multipole wiggler beamline MAGS,~\cite{dudzik06} operated by the HMI at the
synchrotron source BESSY in Berlin. The measurements were carried out in vertical scattering four-circle geometry at a photon energy of 12.398
keV. All $T$ dependencies were measured on heating in order to avoid hysteresis effects, with cooling and subsequent heating rates of $\sim$
2~K/min for these investigations. To avoid sample heating as well as irradiation damages~\cite{blundell06} the power of the beam (full beam
intensity 10$^{12}$ photons/s at 10~keV) was reduced by four orders of magnitude using absorber foils.

For the structure refinement the crystal was mounted to a displex cryostat on a four-circle Huber goniostat using Mo K$_\alpha$ radiation
($\lambda$ = 0.71073 \AA ) and a scintillation counter. Details of the refinement procedure are given below. For these x-ray diffraction
investigations a high cooling rate of $\sim$ 4~K/min was used. According to Ref.~\cite{muller02} this may influence the glass-like transition by
shifting the value of $T_g$ by about 2~K to higher temperatures. As well, if the scenario of a configurational freezing-out is correct, a high
cooling rate should lead to a comparatively high level of residual ethylene group disorder.

\begin{figure}
\begin{center}
\includegraphics[width=0.75\columnwidth]{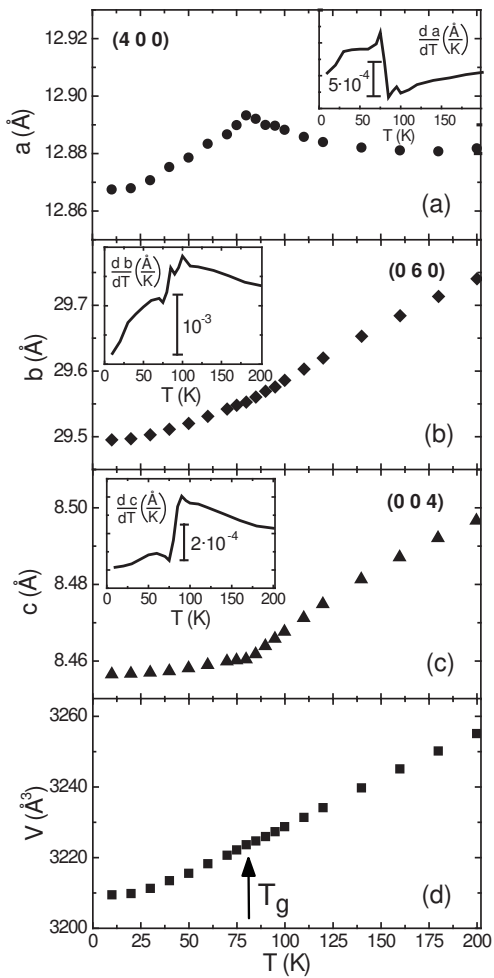}
\end{center}
\caption[2]{(a)-(d) The $T$ dependence of the lattice parameters $a$, $b$ (out of plane), $c$ and the unit cell volume $V$ of $\kappa$-Br from
synchrotron x-ray diffraction. In the insets the derivatives $dx/dT$ of the lattice parameters $x = a, b, c$ are shown.}\label{fig:synchrotron}
\end{figure}

The $T$ dependence of the lattice parameters $a$, $b$, $c$ and the unit cell volume $V$ were investigated  between 10 and 200~K using
high-resolution synchrotron x-ray diffraction. The results of the (4 0 0), (0 6 0) and (0 0 4) Bragg reflections investigated here for the $a$,
$b$ and $c$ parameters, respectively, are depicted in Fig.~\ref{fig:synchrotron}. The data for the three principal axes show a highly
anisotropic behavior. Upon heating, for the $a$ axis an upturn is observed. After reaching its maximum at $T_g$, the lattice constant abruptly
decreases again up to $\sim$200~K. In contrast, for the $b$ and $c$ directions we find a monotonic increase up to 200~K, with only a small kink
at 80~K. Here, $T_g$ is more accurately determined from the temperature derivatives of the lattice parameters (insets of
Fig.~\ref{fig:synchrotron}). However, as result of compensating lattice parameter anomalies, no anomaly in the unit cell volume $V (T)$ is
observed at $T_g$. Our data reproduce the findings on the $T$ dependence of the lattice parameters previously obtained from thermal expansion
and x-ray diffraction~\cite{muller02,watanabe91,kund93,toyota93}.

The anomalous behavior of the $a$ lattice parameter suggests an anomalous change in the bond angle of --CN--Cu--NC-- in the polymeric chain in
the temperature range well above $T_g$. This in turn might induce a change in the hydrogen bonding \mbox{--CH$\cdot\cdot\cdot$N--} between
terminal ethylene groups and nitrogen atoms on the polymer, thus affecting the electronic properties of $\kappa$-Br.

It has been proposed that the anomalies in the $T$ dependence of the lattice parameters are due to a conformational order-disorder transition of
the terminal ethylenes~\cite{kund93}. In order to test this hypothesis, we carried out a detailed study of the temperature dependence of the
atomic coordinates and the occupation factor, OF, of the eclipsed configuration of the (CH$_2$)$_2$ groups. Intensities of 5398, 7045 and 6986
reflections were measured in the range $2^\circ \leq 2 \theta \leq 34^\circ$ for 300, 100 and 9 K, respectively. The lattice constants were
refined by least-square fits to the positions of 72 reflections in the range $25^\circ \leq 2 \theta \leq 30^\circ$. After absorption
corrections the intensity data with $I / \sigma(I) \geq$ 10 were merged to 908, 980 and 965 unique reflections for 300, 100 and 9 K,
respectively. The structure has been refined using the program XTAL~3.4~\cite{xtal95}, and the atomic coordinates and the anisotropic
displacement parameters for non-hydrogen atoms were refined against $|F|$.

One of our central issues was to determine if the crystal symmetry remains the same upon cooling through $T_g$. Systematic absences of
reflections correspond to space group {\it Pnma} for all three temperatures 300, 100 and 9 K, excluding a change of symmetry. Overall, we
conclude that the crystallographic structure of $\kappa$-Br remains essentially the same above and below $T_g$. The lattice parameters, unit
cell volume and fit details of our refinements are summarized in Table~\ref{tab:xray}. Within experimental accuracy, a comparison to data taken
previously yields good agreement regarding lattice parameters, unit cell volume and atomic positions~\cite{strack05,geiser91,geiser91a,kini90}.

\begin{table}
\begin{tabular}{|l|c|c|c|}
\hline
 & 300 K & 100 K & 9 K  \\
\hline\hline
$a$ (\AA )& 12.942(4) & 12.884(5) & 12.869(4) \\
\hline
$b$ (\AA ) & 29.986(7) & 29.581(8) & 29.495(6) \\
\hline
$c$ (\AA ) & 8.542(4) & 8.484(4) & 8.466(3) \\
\hline
$V$ (\AA$^3$) & 3315(3) & 3233(4) & 3214(2) \\
\hline
OF eclipsed config. & 67$\pm$2 \% & 89$\pm$3 \% & 97$\pm$3 \% \\
\hline
unique reflections & 908 & 980 & 965\\
\hline
refined variables & 187 & 187 & 187 \\
\hline
$R$-factor & 0.081 & 0.109 & 0.114 \\
\hline
$wR$-factor & 0.050 & 0.029 & 0.052\\
\hline
\end{tabular}
\caption{Lattice parameters, unit cell volume, occupation factor OF and refinement parameters for $\kappa$-Br, crystallizing in space group {\it
Pnma}, at 300, 100 and 9 K obtained via laboratory x-ray diffraction experiments.}~\label{tab:xray}
\end{table}

\begin{figure}
\begin{center}
\includegraphics[width=0.85\columnwidth]{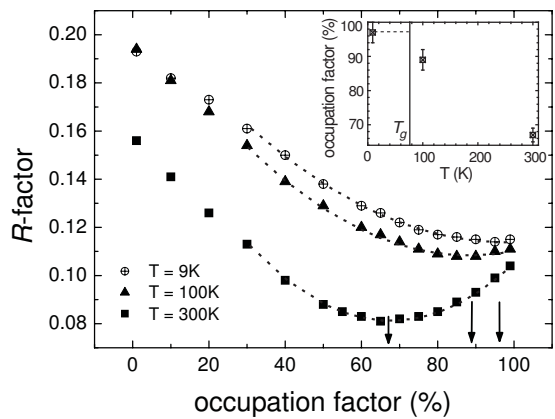}
\end{center}
\caption[2]{The refinement parameter $R$ as function of the occupation factor OF of the (CH$_2$)$_2$ groups in eclipsed configuration at 9, 100
and 300 K. The arrows indicate the positions of minimal $R$-values for 300~K, 100~K and 9~K. The inset depicts the $T$ dependence of OF; for
details see text.} \label{fig:etoccupation}
\end{figure}

The most noteworthy effect of temperature is the tendency towards ordering of the terminal ethylene groups. The values OF were determined using
an iterative refinement procedure. First, a full refinement of 187 variables, {\it i.e.}, atomic coordinates and anisotropic thermal
displacement parameters of all atoms except for terminal ethylene groups, was repeated for different population parameters of the carbon atoms
in eclipsed and staggered configuration in the interval [1$\%$,99$\%$]. Then, using the solution with minimal $R$-factor, a refinement of all
atomic coordinates and thermal parameters~\cite{note3} was performed, iteratively followed by the refinement of all 187 variables and tuning of
the population factor. The results obtained for the last cycle are shown in Fig.~\ref{fig:etoccupation} for temperatures 300, 100 and 9~K. Fits
to the data using a parabolic relation yield an occupation factor at room temperature (100 K) of OF = 67$\pm$2 \% (89$\pm$3 \%) of the ethylene
groups in the eclipsed configuration, in agreement with Ref.~\cite{strack05}. At 9~K, the occupation of the eclipsed configuration is clearly
favorable with a refined value of OF = 97$\pm$3 \%.

As expected for the excitation from eclipsed to staggered configuration involving an energy barrier, the occupation factor of the eclipsed
configuration shows a pronounced $T$ dependence. However, the number of terminal groups in the staggered configuration turns out to be
unexpectedly small with $\sim$11 \% at 100~K, and a residual value of only 3 $\pm$ 3 \% at zero temperature (inset Fig.~\ref{fig:etoccupation}).
This value is the more surprising since we had deliberately chosen relatively high cooling rates in our experiment, which should result in
larger residual disorder values. For most experiments reported in literature, the residual disorder levels thus ought to be even smaller. It
appears highly unlikely that such small residual disorder is responsible for the transition at 77~K. In structural terms, taking into account
that there are 8 pairs of ethylene groups per unit cell, our value at 100~K implies that $\leq$ 8 \% ethylene group disorder leads to an average
distance between staggered end groups of the order of 10-30 \AA \,close to $T_g$, and which increases substantially as temperature is lowered.
In view of the sharp anomalies in thermal expansion measurements~\cite{muller02} as well as the steep steplike increase of the specific heat
upon heating through $T_g$~\cite{akutsu00}, there seems to be some cooperativity between the ethylene moieties at the transition. For a
disorder-order transition or a configurational freezing-out of the ethylene groups to occur under these conditions a structural element
providing a coupling between end groups so far apart is required, that is, the transition at 77~K cannot involve only a freezing-out of terminal
ethylene groups activation. In this context, we note that as a result of our refinement we have obtained unusually large thermal displacement
parameters $U_{ij}$ at low temperatures, {\it i.e.}, $U_{ij}(300 {\rm K})/U_{ij}(100 {\rm K}) \approx U_{ij}(100 {\rm K})/U_{ij}(9 {\rm K})
\approx$ 1.2-1.5, and pronounced anisotropic thermal displacement parameters in particular for the C and N atoms on the polymeric chain. In
accordance with only small changes in the resistivity~\cite{strack05} at $T_g$ despite the pronounced $a$-axis anomaly, this observation might
indicate that the glass-like transition involves reorientational behavior of the insulating polymeric chain rather than the conducting BEDT-TTF
molecule. Unfortunately, even with our detailed structural investigation the experimental resolution is not sufficient to clearly resolve this
issue.

Taking together the requirement for another structural element involved in the glass-like transition and the anomalous $T$ dependence of the
$a$-axis parameter, it is likely that the orbital overlap within the BEDT-TTF molecules is affected. Here, structural transformations of the
polymeric chain will translate into changes of the orbital overlap within the BEDT-TTF layers. This for instance would account for the
experimental observation of a pronounced cooling-rate dependence of the resistivity~\cite{stalcup99}, and which would be difficult to explain
with the small residual level of disorder of terminal ethylene groups in $\kappa$-Br determined experimentally.

In conclusion, for the first time we have directly established the level of residual ethylene group disorder in the organic superconductor
$\kappa$-(BEDT-TTF)$_2$Cu[N(CN)$_2$]Br. We conclude that the glass-like transition does not primarily involve a configurational freezing-out of
the terminal ethylene groups. In consequence, other structural or electronic degrees of freedom must be considered in order to resolve the issue
of relevance of disorder and intrinsic sample properties as for instance an anomalous change in the bond angle in the polymeric chain.

The authors thank J. M\"{u}ller and P. Lunkenheimer for fruitful discussions. This work has been supported by the DFG under contract no.
SU229/8-1. Construction of the beamline MAGS has been funded by the BMBF via the HGF-Vernetzungsfonds under Contract Nos. 01SF0005 and 01SF0006.


\begin{references}
\bibitem{note3} With the small number of carbon atoms in staggered configuration at low $T$, we adopted their atomic coordinates and thermal displacement parameters according to results at 127~K reported in Refs.~\cite{geiser91,geiser91a}.
\end{references}
\end{document}